# Virtual screening of Microalgal compounds as potential inhibitors of Type 2 Human Transmembrane serine protease (TMPRSS2)

**Ibrahim Ahmed Mohammed** [1]

1 Biotechnology Program, School of Bioresources and Technology, King Mongkut's University of Technology Thonburi (Bang Khun Thian), 49 Soi Thian Thale 25, Bang Khun Thian Chai Thale Rd., Tha Kham, Bang Khun Thian, Bangkok, 10150, Thailand.

* Correspondence: ibrahim.amohammed@mail.kmutt.ac.th

## Abstract.

More than 198 million cases of severe acute respiratory syndrome coronavirus 2 (SARS-CoV-2) has been reported that result in no fewer than 4.2 million deaths globally. The rapid spread of the disease coupled with the lack of specific registered drugs for its treatment pose a great challenge that necessitate the development of therapeutic agents from a variety of sources. In this study, we employed an in-silico method to screen natural compounds with a view to identify inhibitors of the human transmembrane protease serine type 2 (TMPRSS2). The activity of this enzyme is essential for viral access into the host cells via angiotensin-converting enzyme 2 (ACE-2). Inhibiting the activity of this enzyme is therefore highly crucial for preventing viral fusion with ACE-2 thus shielding SARS-CoV-2 infectivity. 3D model of TMPRSS2 was constructed using I-TASSER, refined by GalaxyRefine, validated by Ramachandran plot server and overall model quality was checked by ProSA. 95 natural compounds from microalgae were virtually screened against the modeled protein that led to the identification 17 best leads capable of binding to TMPRSS2 with a good binding score comparable, greater or a bit lower than that of the standard inhibitor (camostat). Physicochemical properties, ADME (absorption, distribution, metabolism, excretion) and toxicity analysis revealed top 4 compounds including the reference drug with good pharmacokinetic and pharmacodynamic profiles. These compounds bind to the same pocket of the protein with a binding energy of -7.8 kcal/mol, -7.6 kcal/mol, -7.4 kcal/mol and -7.4 kcal/mol each for camostat, apigenin, catechin and epicatechin respectively. This study shed light on the potential of microalgal compounds against SARS-CoV-2. In vivo and invitro studies are required to developed SARS-CoV-2 drugs based on the structures of the compounds identified in this study.

# 1. Introduction

Corona virus disease-19 (Covid-19) is caused by Severe Acute Respiratory Syndrome Corona Virus 2 (SARSCoV-2) [1]. The disease stamped by pneumonia, feverish condition, dyspnea, fatigue, acute respiratory syndrome (ARDs), multi-organ dysfunction and sputum production was first identified on 31 December 2019 in Wuhan, Hubei, China [2], [3]. The virus was isolated in confirmed pneumonia patients and characterized by polymerase chain reaction and next-generation sequencing techniques [4]. Transmission of the disease in humans is primarily achieved through saliva, nasal discharge, cough or sneezing of the infected person [5]. Infections quite related to Covid-19 caused by Severe Acute Respiratory Syndrome Corona virus (SARS-CoV) and Middle East Respiratory Syndrome Corona Virus (MERS-CoV) were announced in 2003 and 2012 respectively [6]. However, the infectivity of SARSCoV-2 is more pronounced than the later diseases [7] [8]. The expeditious spread of Covid-19 across the globe makes it pandemic and caused a lot of apprehension due to associated mortality rate [9] . According to the World Health Organization (WHO) report of August 2, 2021, SARSCoV-2 has affected 223 countries with 198,385,991 confirmed cases that result in the death of more than 4.2 million people world-wide [WHO]. Consequent upon the emergence of the pandemic, there was a global economic crisis that led to the colossal downturn in production, loss of jobs, and restriction of social movement [10].

Coronaviruses are spherically shaped protein-coated viruses with single stranded, positive sense RNA genome surrounded by a spike glycoprotein that is critical for viral attachment and entry into the host cell [11]. Their genome is the largest of all known RNA viruses ranging from 20-32 kilo base pairs that encodes different proteins involved in viral life cycle [12][8]. There are four structural proteins (SPs) and about sixteen non-structural proteins (NSPs) encoded by coronaviruses genome [13]. The structural proteins are; spike (S) protein which mediate viral interaction with host cell Angiotensin Converting Enzyme receptor type 2 (ACE-2), Matrix (M) protein that is responsible for the assembly of virions and their entrapment in between Endoplasmic reticulum and Golgi apparatus for their fusion into the newly produced virions [14][15]. Other structural proteins are Nucleocapsid (N) which assist in the processing of viral genome and Envelop protein (E) for viral proliferation and development [16]. These proteins are vital hotspots for inhibiting the viral proliferation and consequently the spread of SARSCoV-2 infection.

Bioactive compounds (BCs) from different natural sources were successfully developed and used in the treatment of several viral diseases. Microalgae are natural reservoirs of awfully fascinating active compounds. These compounds are exceedingly remarkable and have gained wider scientific attention due to their broad distribution, numerous functions and captivating properties [17]. Colossal number of studies have reported production of collection of chemical compounds by microalgae with contrasting bioactivities such as phycobilins, polyunsaturated fatty acids, vitamins, polysaccharides, proteins, and sterols among others. Numerous microalgal products with antiviral, antimicrobial, anticancer, antifungal, and antioxidant activities were established [17]. *Chlorella pyrenoidosa* extract containing acid polysaccharide demonstrates inhibitory activity against vesicular stomatitis virus (VSV) in mice [18]. Compounds from cyanobacteria such as *lyngbya* and green alga like *Dunaliella primolecta* were found to be active inhibitors of Herpes Simplex Virus type 1 (HSV-1) [19]. Astaxanthin from *Haematococcus pluvialis* and some diatoms displayed activity against white spot syndrome virus (WSSV). This compound also demonstrates anticancer property by up regulating the expression of p53 and some cyclin kinase inhibitors and decreasing the expression of cyclin D1 in colon cancer cells [19], [20]. In vivo and in vitro studies have shown inhibitory effect of *A. platensis* compounds against Influenza virus [21], Herpes Simplex Virus type 1 (HSV-1) [22] and Human immunodeficiency virus type -1 (HIV-1) [23] among others. Carotenoids extract from *D. salina* and *H. pluvialis* significantly inhibited the activity of HSV-1 [24].

Research towards the use of microalgal compounds as drugs candidates against SARSCoV-2 is limiting and highly needed. One of the most important SARSCoV-2 drugs target protein is type-2 human transmembrane serine protease (TMPRSS2) which prime S protein for viral entry to the host cell via ACE-2 receptor [1]. Targeting this protein is highly critical in halting the viral pathogenicity as it is found in human parts more prone to infection such as digestive tract, airways, cardiac endothelium, kidney and microvascular endothelial cells [25]. In this research, an in silico virtual screening of 95 microalgal active compounds was carried out to identify the most active hits capable of binding and inhibiting human TMPRSS2. ADME and Toxicity prediction analysis was conducted to further determined drug-likeness, pharmacokinetic and pharmacodynamic parameters of the active hits leading to the identification of top 4 leads. Camostat; a known commercial TMPRSS2 inhibitor drug was used as reference ligand in each predictive step. This study has explored the potential of microalgal active compound as a promising drug candidate against TMPRSS2 that could be develop and use as drugs in the treatment of Covid-19.

## 2. Materials and Method

### 2.1 TMPRSS2 homology modeling

The three-dimensional structure of TMPRSS2 is not readily available in protein data bank (PDB) server, we therefore sought to construct the model of this protein. The protein contains 492 amino acids residues whose sequence (**Figure 1**) were obtained from UniProtKB database [26] in FASTA format. The UniProtKB identifier of the TMPRSS2 is O15393-1. The amino acid sequences were uploaded in Iterative Threading ASSembly Refinement (I-TASSER) [27], an online web server used for the prediction of protein structure and function for the construction of 3D model of TMPRSS2.

### 2.2 Protein model refinement and validation

Protein model refinement is highly important to improve its structural quality. The 3D structure of the protein's model was refined using GalaxyRefine web server [28]. GalaxyRefine is a robust online server widely used for the enhancement of both local and global quality of protein structure via utilization of excellent protein structure forecasting web tool. The top five (5) refined protein models were obtained from GalaxyRefine, and the best model was selected. PyMOL (https://pymol.org/2/), was used to visualized and label different domains of the protein model (**Figure 2**). The quality of the model was validated using Ramachandran plot server [29] which produces Ramachandran plot that enable visualization of highly preferred, preferred, and disallowed phi (φ) and psi (ψ) angles of each amino acid in a protein. The overall quality of the chosen model was checked with the aid of Protein Structure Analysis (ProSA) web tool [30], to ascertain the overall quality of the protein. This server generates a score for a given input protein known as the Z-score. For a high-quality protein model, Z-score lies within the range of native proteins otherwise the protein structure might have some errors.

```
>sp|O15393|TMPS2_HUMAN Transmembrane protease serine 2 OS=Homo sapiens
OX=9606 GN=TMPRSS2 PE=1 SV=3

MALNSGSPPAIGPYYENHGYQPENPYPAQPTVVPTVYEVHPAQYYPSPVPQYAPRVLTQA
SNPVVCTQPKSPSGTVCTSKTKKALCITLTLGTFLVGAALAAGLLWKFMGSKCSNSGIEC
DSSGTCINPSNWCDGVSHCPGGEDENRCVRLYGPNFILQVYSSQRKSWHPVCQDDWNENY
GRAACRDMGYKNNFYSSQGIVDDSGSTSFMKLNTSAGNVDIYKKLYHSDACSSKAVVSLR
CIACGVNLNSSRQSRIVGGESALPGAWPWQVSLHVQNVHVCGGSIITPEWIVTAAHCVEK
PLNNPWHWTAFAGILRQSFMFYGAGYQVEKVISHPNYDSKTKNNDIALMKLQKPLTFNDL
VKPVCLPNPGMMLQPEQLCWISGWGATEEKGKTSEVLNAAKVLLIETQRCNSRYVYDNLI
TPAMICAGFLQGNVDSCQGDSGGPLVTSKNNIWWLIGDTSWGSGCAKAYRPGVYGNVMVF
TDWIYRQMRADG
```

**Figure 1**. amino acids sequence of TMPRSS2 showing N-terminal LDL-receptor class A domain (113-148 blue amino acid residues), Scavenger Receptor Cysteine Rich domain (SRCR) (153-246 hot pink amino acid residues) and C-terminal peptidase S1 domain (256-487 cyan amino acid residues).

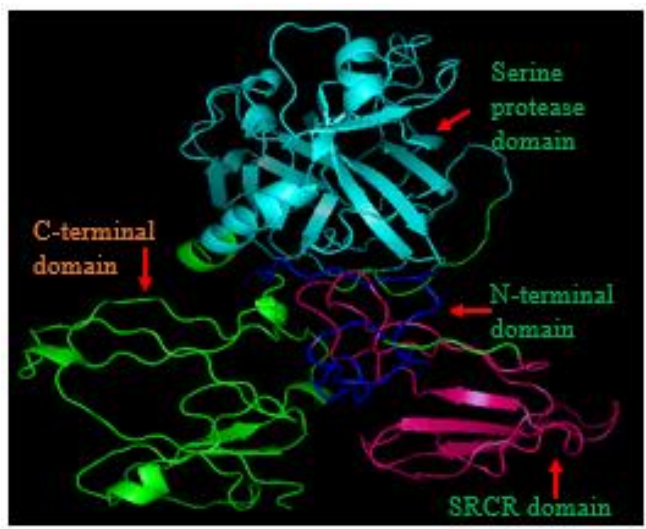

**Figure 2**. 3D structure of TMPRSS2 developed through homology modeling displaying N-terminal domain (blue), Scavenger Receptor Cysteine Rich domain (SRCR), C-terminal domain and Seine proteases domain

### 2.2 Ligands selection and retrieval

A literature survey on bioactivities of different microalgal compounds was carried out that led to the identification of 95 active compounds. The compounds were selected based on their reported activities such as antibacterial, antiviral, anticancer and anti-inflammatory activities. 3D-structure of all the ligands and the reference standard compound (Camostat) were downloaded from PubChem database (https://pubchem.ncbi.nlm.nih.gov) and saved in structure data format (sdf). PyRx software [31] was used for the virtual screening of the ligands.

## 2.3 Ligands Binding Site Prediction

Inhibition of proteins by a small organic molecule is a function of protein-ligands interaction which is achieved at a defined protein binding site. It is therefore highly critical to determine the ligand binding pocket to enable proper elucidation of the nature of inhibition, amino acids involved and kind of interaction. Ligands binding site and the residues of TMPRSS2 were predicted using COACH-D server [32]. The pdb file of the protein was inputted to the server which then employs five-steps algorithm that exploit ligand-binding site of a protein. This is achieved through identification of a ligand-binding templates from BioLiP database of annotated ligand-protein complexes [33]. The refined binding pose is then recognized by efficient docking of the template or the query ligand into the query protein. The top five predicted ligand binding poses are generated as the output and ranked according to the parameters such as confidence score (C-score) and the binding cluster size. Out of the top five predicted binding poses of TMPRSS2, we selected the best pose based on the C-score and binding cluster size. Ligand binding site prediction was further complimented using 3DLigandSite server (ic.ac.uk). This web tool operates based on the predicted homologous proteins from the library of proteins structures with bound ligands, to project ligand binding site of a query protein. Protein data bank file of TMPRSS2 model was submitted to the server which search for ligand-bound homologous proteins and their subsequent alignment to the query protein. Ligand clustering was then used to predict the protein probable binding pocket. The cluster containing the highest number of ligands is considered as the binding site.

## 2.3 Virtual screening/ molecular docking

To carryout virtual screening of the selected compounds, the TMPRSS2 protein model was loaded into the PyRx virtual screening tool in .pdb format. PyRx is a virtual screening software for computational drugs discovery that operate based on empirical-based free energy scoring function and Lamarckian Genetic Algorithm (LGA). The protein was converted to .pdbqt by addition of polar hydrogens and Kollman charges. The 95 ligands in .sdf format were manually inputted and energy-minimized using universal force field of PyRx software and finally converted to. pdbqt format. The grid box that specifically mapped the predicted binding domain (serine protease domain) of TMPRSS2 was created. The grid centers used were x = 75.18, y = 91.08, and z = 66.32 with dimension of 38.80 x 40.04 x 38.10 Å. The compounds were virtually screened / docked compounds against TMPRSS using PyRx in-build Autodock vina [31], [34]. Best seventeen (17) hits were generated which were then further subjected to ADME (absorption, distribution, metabolism, excretion) and Toxicity analysis to finally obtained best four (4) leads. The PubChem ID and name of the best 17 hits are presented in **table 2** with the control compound (camostat) highlighted in bold. ACD/ChemSketch (ACD/labs 2020.2.1) was used to draw the two-dimensional structure of the most active compounds (**Figure 3**).

## 2.4 Analysis of protein-ligands interaction

The interaction of the compounds with TMPRSS2 was carried out using graphic user interface of Autodock tool (version 1.5.6) [35], Discovery Studio 2020 and LigPlot$^+$ (v.2.2.4) [36]. The surface view of the protein was created with Autodock tool while 2D-interaction analysis was performed using Discovery studio and LigPlot$^+$

## 2.5 ADME Analysis.

A crucial step in drug discovery is the analysis of pharmacological activity of drug candidate. This process is labor intensive and time demanding under laboratory setting. In silico prediction of pharmacophore properties of the active compounds is therefore important in accelerating the process of drug development. The properties are absorption, distribution, metabolism, excretion, and toxicity (ADME and T). These parameters provide information on the possibility of placing an active compound into living subject. Analysis of ADME was performed by submitting canonical simplified molecular input line entry system (SMILES) of the compounds to an online server; ADMETlab 2.0 [37].

## 2.6 Toxicity studies

In order to ascertain and verify the safety of the selected compounds for human consumption, we performed toxicity prediction studies using ProTox-II [38]; a virtual web tool employed for the assessment of toxicity of compounds. The compounds names were inputted to the server and selected toxicity prediction were chosen which includes lethal dose 50 ($LD_{50}$), hepatoxicity, carcinogenicity, and cytotoxicity. All the parameters were compared with the commercially known TMPRSS2 inhibitor drug; Camostat.

**Table 2**. PubChem ID and name of the best hits used in this study.

| Compound | PubChem CID | Compound | PubChem CID |
|---|---|---|---|
| **Camostat** | **2356** | Comnostin B | 10093982 |
| Cybastacine A | 139051842 | Quercetin 3-galactoside | 5281643 |
| Calothrixin A | 9926867 | Cryptophycin 1 | 6438401 |
| Scytoscalarol | 44605340 | Quercetin | 5280343 |
| Quercetin 3-rutinoside | 5280805 | Comnostin E | 10025501 |
| Noscomin | 10410225 | Bauerine C | 189883 |
| Apigenein | 5280443 | 4-phenylacridine | 620325 |
| Catechin | 9064 | Comnostin D | 10390076 |
| Epicatechin | 72276 | | |

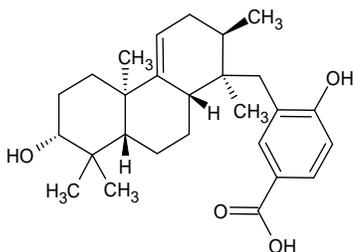
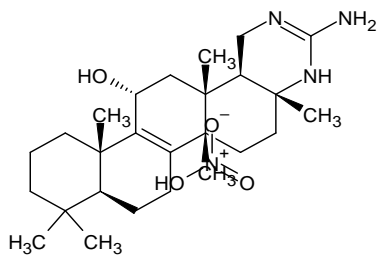
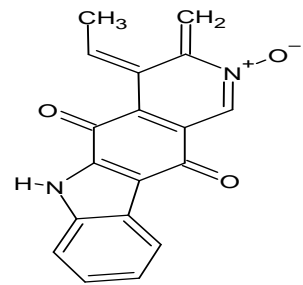

(a) Noscomin  (b) Cybastacine A  (c) Calothrixin A

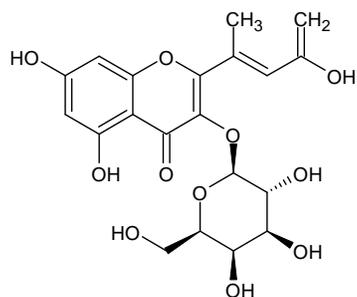 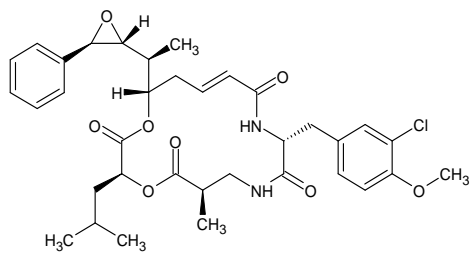 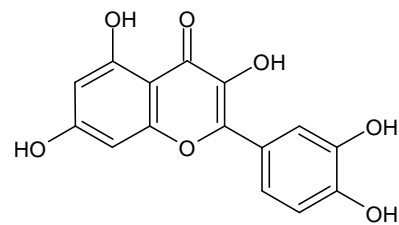

**(j) Quercetin 3-galactoside**   **(k) Cryptophycin 1**   **(l) Quercetin**

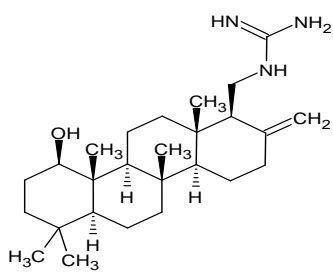 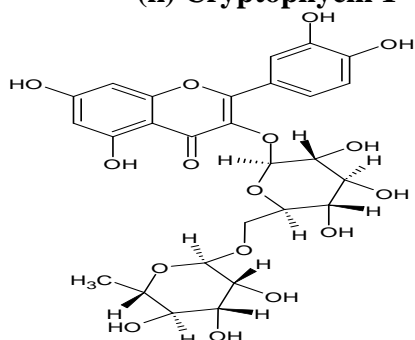 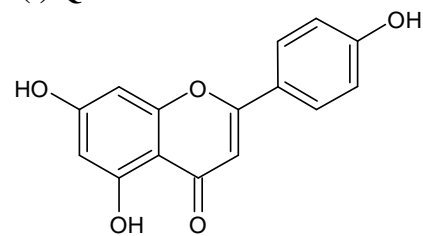

**(a) Scytoscalarol**   **(e) Quercetin 3-rutinoside**   **(f) Apigenin**

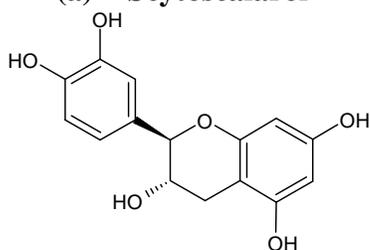 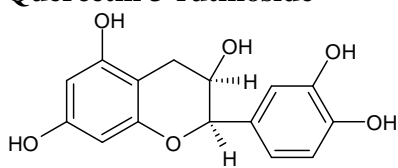 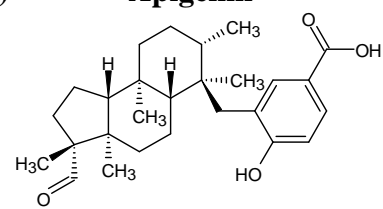

**(g) Catechin**   **(h) Epicatechin**   **(i) Comnostin B**

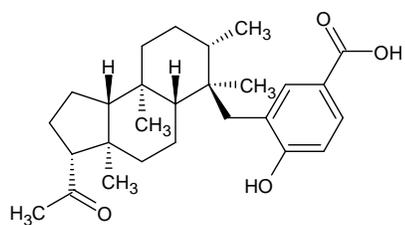 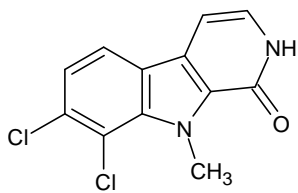 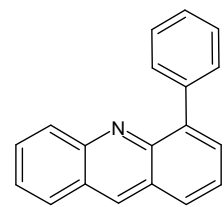

**(m) Comnostin E**   **(n) Bauerine C**   **(o) 4-phenylacridine**

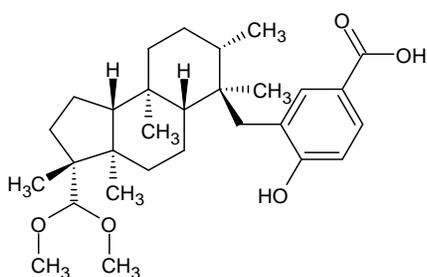 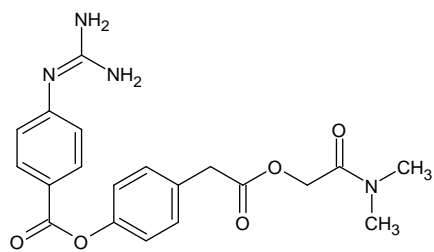

**(p) Comnostin D**   **(q) Camostat**

**Figure 3.** Two-dimensional structures of top 17 microalgal compounds (**a-q**) based on their affinity to bind and inhibit TMPRSS2

## 3. Result and Discussion.

### 3.1 TMPRSS2 homology modeling

The model of TMPRSS2 was constructed by I-TASSER which predict 3D-structure of a protein model by excising continuous fragments from threading alignment and subsequently reassembling them using replica-exchanged Monte Carlo simulation. The software generates the top five models which are evaluated based on some parameters such as C-score, TM-score and RMSD and cluster density. C-score indicates how confidence the quality of the predicted protein model is. The score is rated between -5 to 2 with higher score signifying good quality model and vice-versa. TM-score and RSMD are used to indicate the structural similarity between the query and template proteins. TM-score of >0.5 indicate model with correct structure and topology while <0.17 means that there is random similarity between the query and template proteins. Cluster densities reflect the frequency of the occurrence of protein structure in the simulated trajectory, with higher value denoting better quality model. We selected model 1(**Figure 4 a**) which was considered as the best having C-score of -0.33, TM-score of 0.67±0.13 and cluster density of 0.2273. Both parameters are quite adequate thus implying the quality and acceptability of our protein model.

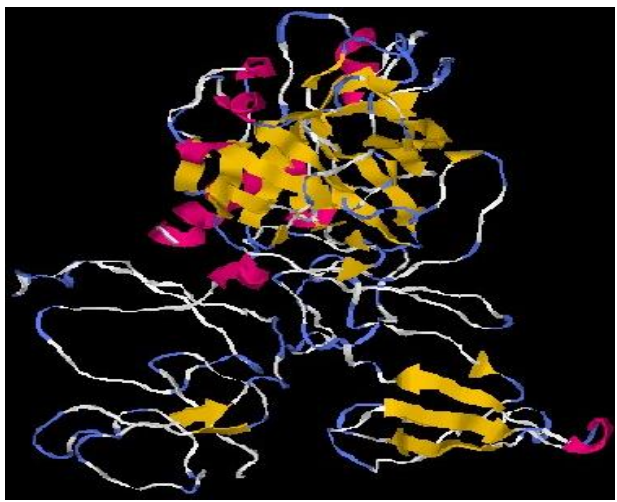

a

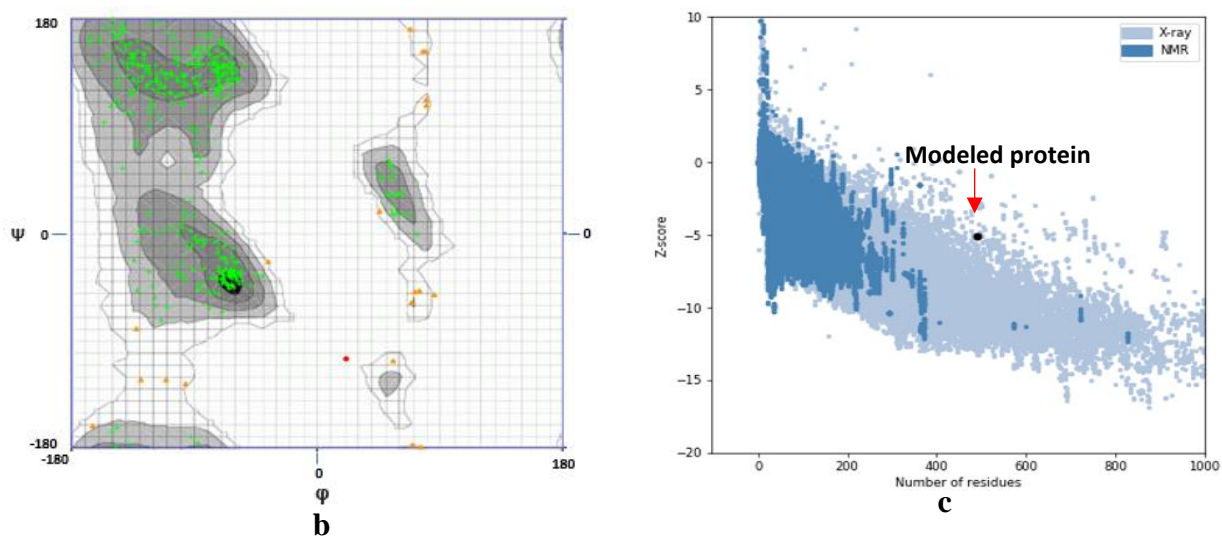

**Figure 4** structure and validation of TMPRSS2 model. (**a**) 3D- structure of the refined protein model, (**b**) Ramachandran plot of the refined model showing 95.169 % (green), 4.589.072 % (orange), and 0.242 % (red) of the amino acids residues in the highly preferred, preferred, and disallowed regions respectively, (**c**) Final protein model quality with Z-score of -5.1

### 3.2 Protein model refinement and validation

The 3D-model of TMPRSS2 (**Figure 4 a**) was refined with the aid of GalaxyRefine web server which generated five models. The best model was selected based on GalaxyRefine parameters which are GDT-HA, RMSD, Molprobity, clash score, poor rotamers and Rama favored. Consequently, model 1 was considered the best which has 89.6 % residues in the favored region, RSMD of 0.472, GDT-HA of 0.9370, clash score of 15.1 and 0.5 poor rotamers. The quality of the refined model was validated by Ramachandran plot server, unveiling 95.169 % (green), 4.589.072 % (orange), and 0.242 % (red) of the amino acid residues in the highly preferred, preferred, and disallowed regions respectively (**Figure 4 b**). The validated model of a protein may sometimes present a problem that needs to be detected and address to obtain a high integrity, consistent, and reliable model. In this regards, ProSA web tool was employed to check for the potential errors in the validated model. The server display score (Z-score) and energy plot that indicate the position of a problem spotted in the protein structure. In particular, Z-score measure the deviation of proteins structure total energy in relation to the energy distribution obtained from the random conformations [30]. Z-score should normally be within the range of a characteristic native proteins' structures obtained through x-ray and NMR, otherwise the model protein structure is erroneous. Interestingly, the Z-score obtained for our model is -5.1 which lies within the range of most of native proteins (**Figure 4 c**).

### 3.3 Ligands binding site prediction

An important aspect in structure-based drug design (SB-DD) is the prediction and analysis of ligand binding site. This is because molecular association takes place at this site of whose knowledge is crucial to decipher biological activities of a protein. Understanding the binding pocket can also be used to provide an insight on the development of treatment strategy of diseases such as Covid-19. Computer aided binding site prediction is widely used in computational drugs design and discovery which offer advantage of saving time and cost associated with laboratory-based approach. The modelled TMPRSS2 binding site was predicted using COACH-D server and further verified by 3DLigandSite server (ic.ac.uk).

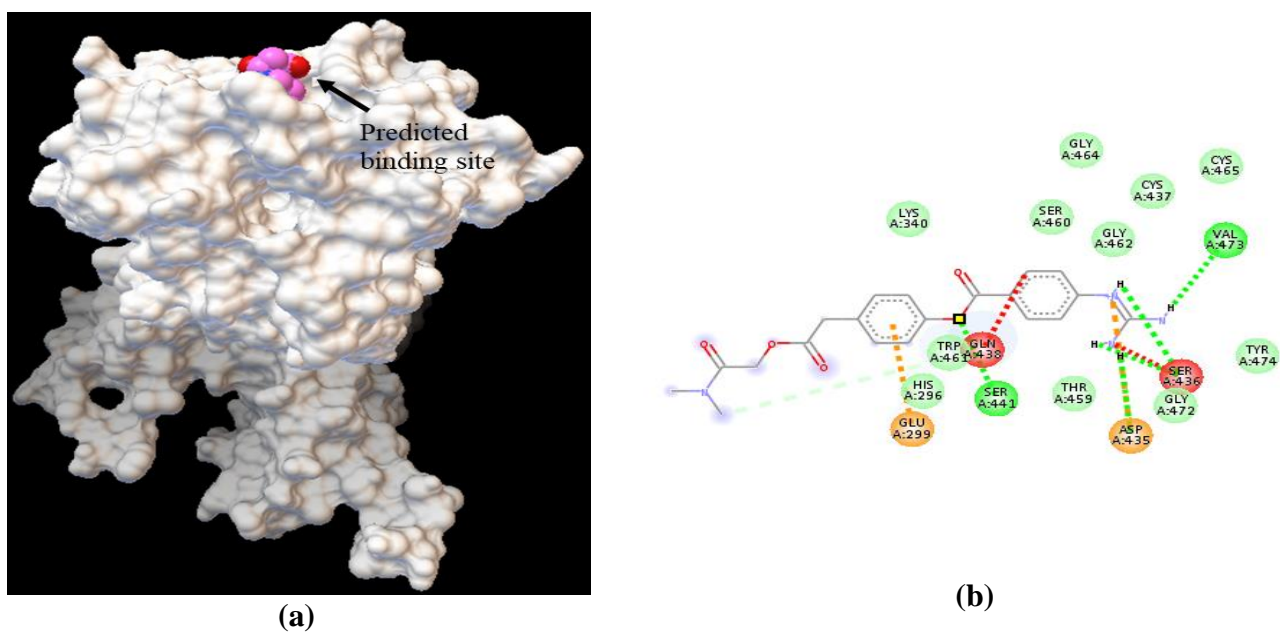

(a) (b)

**Figure 5.** Ligand binding site prediction of TMPRSS2 (**a**) 3D-predicted binding site (**b**) 2D-interaction of the ligand with amino acids residues of TMPRSS2

The best predicted binding pocket as revealed by COACH-D server was selected based on C-score of 0.61 and cluster size of 1107 which are both quite good compared to the remaining predicted sites, thus implying a high reliability of the selected pose. The forecasted binding pocket (**Figure 5 a**) was detected at the serine protease domain of the protein that contain the catalytic triad and substrate binding site. The catalytic triad of TMPRSS2 consist of His296, Asp345 and Ser441 while the substrate binding residues are Asp435, Ser460 and Gly462 [33]. Interestingly, the predicted ligand pocket of TMPRSS2 is surrounded by both the substrate binding residues catalytic triad residues except Asp345 (**Figure 5b**), thus indicating the efficiency and reliability of our prediction. Other amino acids residues found at the vicinity of TMPRSS2 binding pose are Glu299, Trp461, Gln438, Thr459, Gly472, Ser436, Tyr474, Val473, Lys340, Cys437, Gly464 and Cys465 (**Figure 5b**). Moreover, 3DLigandSite server generated 9 clusters on TMPRSS2 each with different number of ligands. The cluster containing the highest number of ligands (cluster 1) was selected as the most likely ligand binding site of TMPRSS2. Excitingly, this cluster lies within the serine protease domain of the protein which further validate the earlier prediction based on COACH-D server.

### 3.4 Molecular docking

A total of 95 compounds from different microalgal species were selected based on their bioactivities and docked against TMPRSS2 using in-built Autodock vina of PyRx software. The docking result has revealed 17 compounds with binding energy comparable, a bit higher or lower than that of TMPRSS2 standard inhibitor Camostat, an FDA- approved drug. These compounds have the highest negative binding scores and are considered as most likely inhibitors of TMPRSS2. The binding energies (BEs) of the top 17 compounds is presented in **table 3.**

**Table 3.** Binding energies of the top hits against TMPRSS2

| Compound | Binding energy (kcal/mol) | Compound | Binding energy (kcal/mol) |
|---|---|---|---|
| Camostat | -7.8 | Comnostin B | -7.3 |
| Cybastacine A | -8.1 | Quercetin 3-galactoside | -7.3 |
| Calothrixin A | -7.9 | Cryptophycin 1 | -7.3 |
| Scytoscalarol | -7.8 | Quercetin | -7.2 |
| Quercetin 3-rutinoside | -7.8 | Comnostin E | -7.2 |
| Noscomin | -8.2 | Bauerine C | -7.1 |
| Apigenein | -7.6 | 4-phenylacridine | -7 |
| Catechin | -7.4 | Comnostin D | -7 |
| Epicatechin | -7.4 | | |

The docking result obtained have shown that Camostat, a Japanese FDA-approved drug used for the therapy of chronic pancreatitis, postoperative and reflux esophagitis [39] interacted with TMPRSS2 with a binding energy of -7.8 kcal/mol. This is comparable to the binding energies of Scytoscalarol and quercetin 3-o- rutinoside each with BE of -7.8 kcal/mol, respectively. However, Noscomin, Cybastacine A and Calothrixin A; antibacterial and anticancer compounds found in *Nostoc sp* and *Calothrix sp* respectively [40], displays a better binding affinity against TMPRSS2 with BE of -8.2 kcal/mol, -8.1 kcal/mol and -7.9 kcal/mol, respectively. Other microalgal compounds with good docking scores demonstrate BE in the range of -7.6 kcal/mol to -7.0 kcal/mol. This result implies the existence of active compounds in microalgae that could efficiently bind and inhibit human TMPRSS2 better or comparable to the Camostat and may likely be consider and develop as potential drugs candidate against Covid-19. Drug development entails a series of steps that ensure the safety profile and determination of pharmacokinetic properties of the drug candidate. Thus, binding affinity alone cannot be used to rationalize the applicability of active compounds in the treatment of diseases. In line to this, we performed ADME analysis of the active hits for further screening.

### 3.5 Drug-likeness determination

Most of the problems encountered during drugs development is poor pharmacokinetic properties of drug candidates [37]. Determination of drug-like behavior, absorption, distribution, metabolism, and excretion (ADME) are crucial in the early step of drug development for the identification of the lead compounds. Drug-like properties and ADME analysis of the selected compounds was carried out using ADMETlab 2.0 [37] which revealed six (6) compounds with quite good profile. A perfect drug should be properly absorbed by the body, distributed realistically to various tissues and organs, metabolized efficiently and eliminated appropriately [41]. In this analysis, we first considered physicochemical properties of the hits as molecular weight (MW), hydrogen bond acceptors (HBAs), hydrogen bond donors (HBDs), number of rotatable bonds (RBs), topological surface area (TPSA) or polar surface area, water/octanol partition coefficient (log P) and synthetic accessibility score (SAscore). SAscore is defined as the ease of synthesis of drug-like molecule. These parameters were used to assessed drug-likeness behavior of the compounds based on

Lipinski rule (LR), Pfizer rule (PR), GlaxoSmithKline rule (GSKR) and Golden Triangle rule (GTR). According to Lipinski rule, an ideal drug should have a MW ≤ 500 g/mol, logP ≤ 5, hydrogen bond acceptors ≤ 10 and hydrogen bond donors ≤ 5. PR and GSKR are based on MW, logP and TPSA. Compound's logP and TPSA should be >3 and < 75 respectively in PR while logP and molecular weight should be ≤ 4 and ≤ 400 respectively stipulated in GSKR. In agreement to GTR, MW of a drug should be confined within 200 ≤ MW ≤ 50 and logD (lopgP at physiological PH of 7.4) in the range of -2 ≤ logD ≤ 5. In these studies, the hits that satisfied all these criteria were selected with reference to standard drug; Camostat. Albeit all the compounds could be synthesized without difficulty as justified by their SAscore of less than 6, only 6 of them (**Table 4)**, highlighted in bold) scored all the rules including the reference drug. These are, Calothrixin A, Camostat (reference drug), apigenin, catechin, epicatechin and quercetin. By virtue of their interesting physicochemical properties and incredible drug-likeness behavior and good affinity to TMPRSS2, these compounds may present favorable ADME profile that could be useful to provide an insight on their development as drugs that can be harnessed to inhibit TMPRSS2. The 6 drug-like compounds were then subjected to ADME analysis to further establish their pharmacokinetic properties.

**Table 4.** Physicochemical properties of the selected compounds.

| Compounds | MW | HBA | HBD | RB | TPSA | LOGP | SAscore | LP | PR | GSKR | GTR |
|---|---|---|---|---|---|---|---|---|---|---|---|
| Noscomin | 426.28 | 4 | 3 | 3 | 77.76 | 5.358 | 4.358 | accepted | Accepted | rejected | accepted |
| Cybastacine A | 413.34 | 4 | 4 | 0 | 70.64 | 4.148 | 5.22 | accepted | Rejected | rejected | accepted |
| **Calothrixin A** | **314.07** | **5** | **1** | **0** | **76.87** | **3.11** | **2.79** | **accepted** | **Accepted** | **accepted** | **accepted** |
| Scytoscalarol | 415.36 | 4 | 5 | 2 | 84.63 | 3.985 | 4.861 | accepted | Accepted | rejected | accepted |
| Quercetin 3-rutinoside | 610.15 | 16 | 10 | 6 | 269.43 | -0.763 | 4.783 | accepted | Rejected | rejected | rejected |
| **Camostat** | **398.16** | **9** | **4** | **10** | **137.31** | **1.261** | **2.28** | **accepted** | **Accepted** | **accepted** | **accepted** |
| **Apigenein** | **270.05** | **5** | **3** | **1** | **90.9** | **3.307** | **2.253** | **accepted** | **Accepted** | **accepted** | **accepted** |
| **Catechin** | **290.08** | **6** | **5** | **1** | **110.38** | **1.142** | **3.344** | **accepted** | **Accepted** | **accepted** | **accepted** |
| **Epicatechin** | **290.08** | **6** | **5** | **1** | **110.38** | **1.213** | **3.344** | **accepted** | **Accepted** | **accepted** | **accepted** |
| Comnostin B | 426.28 | 4 | 2 | 4 | 74.6 | 4.688 | 4.484 | accepted | Rejected | rejected | accepted |
| Quercetin 3-galactoside | 464.1 | 12 | 8 | 4 | 210.51 | -0.17 | 4.008 | rejected | Accepted | rejected | accepted |
| Cryptophycin 1 | 634.33 | 10 | 1 | 8 | 132.89 | 4.7 | 5.215 | accepted | Accepted | rejected | rejected |
| **Quercetin** | **302.04** | **7** | **5** | **1** | **131.36** | **2.155** | **2.545** | **accepted** | **Accepted** | **accepted** | **accepted** |
| Comnostin E | 426.68 | 4 | 2 | 4 | 74.6 | 4.961 | 4.27 | accepted | Rejected | rejected | accepted |
| Bauerine C | 266 | 3 | 1 | 0 | 37.79 | 3.197 | 2.621 | accepted | Rejected | accepted | accepted |
| 4-Phenylacridine | 255.1 | 1 | 0 | 1 | 12.89 | 4.927 | 1.665 | accepted | Rejected | rejected | accepted |
| Comnostin D | 472.32 | 5 | 2 | 6 | 75.99 | 5.438 | 4.621 | Accepted | Accepted | rejected | accepted |

Note: **MW**: molecular weight ($\leq 500$); **HBA**: Hydrogen bond acceptors ($\leq 10$); **HBD:** Hydrogen bond donors ($\leq 5$); **RB**: Rotatable bonds (0~11); **TPSA**: Topological surface area ($< 140$); **LOGP**: Water/octanol partition coefficient (0~3); **SAscore**: Synthetic accessibility score (score of $\geq 6$ means the compound is difficult to synthesize while $< 6$ indicate easy to synthesize molecule); **LP**, **PR**, **GSKR** and **GTR** represent Lipinski rules, Pfizer rules, GlaxoSmithKline rules, and Golden triangle rules respectively, (refer to the text for detail).

**Table 5**. Pharmacokinetic profile of the best 6 compounds

| Profile | Compounds | | | | | |
|---|---|---|---|---|---|---|
| | Camostat | Calothrixin A | Apigenin | Catechin | Epicatechin | Quercetin |
| *Absorption* | | | | | | |
| Caco-2 permeability (log unit) | -5.391 | -5.453 | -4.847 | -5.971 | -6.213 | -5.204 |
| Human intestinal absorption | + | + | + | + | + | + |
| P-glycoprotein inhibitor | + | - | - | - | - | - |
| P-glycoprotein substrate | - | - | - | - | - | - |
| *Distribution* | | | | | | |
| Plasma protein binding (%) | 67.96 | 95.41 | 97.25 | 92.06 | 92.35 | 95.49 |
| Volume of distribution (L/kg) | 0.684 | 0.524 | 0.510 | 0.661 | 0.652 | 0.579 |
| BBB penetration | + | + | - | - | - | - |
| *Metabolism* | | | | | | |
| CYP2C9 inhibitor | - | - | + | - | - | - |
| CYP2C9 substrate | + | - | - | - | - | - |
| CYP2D6 inhibitor | - | - | - | - | - | - |
| CYP2D6 substrate | + | - | - | - | + | - |
| CYP3A4 inhibitor | - | - | - | - | - | + |
| CYP3A4 substrate | + | + | - | - | - | + |
| *Excretion* | | | | | | |
| Clearance (ml/min/kg) | 7.163 | 3.594 | 7.022 | 17.911 | 16.512 | 8.284 |
| Half-life (probability of high $T_{1/2}$) | 0.467 | 0.065 | 0.856 | 0.853 | 0.884 | 0.929 |

### 3.6 ADME Analysis

The prediction of ADME profiles via in silico approach has long been established steady in providing a reliable pharmacokinetic property of a compounds. To exert their therapeutic action, orally administered drug should be efficiently absorbed to the blood stream, distributed to various tissues and organs, metabolized by enzymatic system, and finally excreted by the body. From **table 5**, it is evident that all the compounds could be absorbed by the intestinal mucosa as revealed by their positive human intestinal absorption. The absorption of the compounds via blood stream was further disclosed by caco-2 permeability values that range from -4.847 to – 5.971 log unit. In ADMETlab 2.0 server, a permeability of > -5.15 log unit indicate optimal caco-2 absorption. An important determinant of drug absorption is efflux drug transporter; plasma glycoprotein (P-glycoprotein). The substrates of this protein are pumped back to the intestinal lumen thus decreasing their absorption. Furthermore, inhibitors of p-glycoprotein decrease the bioavailability of drugs known to be transported by it. In our analysis, all the compounds except camostat are negative substrates and inhibitors of p-glycoprotein which further explain good absorption profile of compounds under study. Distribution of drug is largely influenced by drug-protein interaction which in turn affect its pharmacokinetic and pharmacodynamic behavior [42]. It is therefore imperative to estimate plasma protein binding as part of characterization of new chemicals during drug development. None of the designated compounds is distributed freely without binding to a plasma protein. This is particularly more noticeable amongst calothrixin A, apigenin, catechin and epicatechin which are more than 90 % protein bound each while only 67.96 % of camostat is in association with protein. Protein-drug interaction also affect the distribution of the drug in the whole body in relation to plasma (volume of distribution). The optimal volume of distribution (VD) of a drug is in the range of 0.04 to 20 L/kg as embedded in the ADMETlab 2.0 server. The compounds under investigation have VD that satisfied these criterions. Interestingly, the VD were comparable amongst all the compounds including camostat, the standard drug, which lies between 0.51 L/kg (apigenin) to 0.684 L/kg (camostat). With exception of camostat and Calothrixin A, the compounds cannot pass blood brain barrier (blood brain barrier negative) and their intake may not be linked to the development of neurological disorders. Therapeutic action of a drugs is also determined by their ability to inhibit or act as a substrate of cytochrome P450 (CYP450) subfamily. This is important to foresee the nature, effectiveness, or harmfulness of co-administration with proven CYP450 substrates [43]. In table 5, camostat is a negative inhibitor and substrate of CYP2C9, CYP2D6, and CYP3A4, respectively. Other compounds in this study are mostly non-inhibitors and non-substrates of CYP450 isoforms except calothrixin A, apigenin and epicatechin which are substrate, inhibitor, and substrate of CYP3A4, CYP2C9 and CYP2D6, respectively. In addition, quercetin is both a substrate and inhibitor of CYP3A4 and non-inhibitor and non-substrate of other isoforms. This information can be used to appropriately develop and place these compounds as drugs that can be used to treat covid-19. Following administration, absorption, distribution and metabolism, the drug and/ or its metabolites should be excreted by the body. This is determined by the drug half-life and renal clearance. A highly, moderately, and slowly excreted drug should have a clearance of > 15 ml/min/kg, 5-15 ml/min/kg and < 5 ml/min/kg, correspondingly. catechin and epicatechin demonstrate a high renal clearance of 17.911 ml/min/kg and 16.512 ml/min/kg, respectively. Camostat, apigenin and quercetin exhibits moderate clearance while calothrixin A show low clearance of 3.594 ml/min/kg. The low clearance of calothrixin A is compensated by its low probability (0.065) of having long half-life thereby preventing possible unwanted effects. This implies that most of these compounds may not pose a threat to the body consequent upon their promising renal clearance.

### 3.7 Toxicity studies

Next we look into the safety profile of the top 6 compounds by performing toxicity prediction studies using online tool; ProTox-II [38]. This server categorized compounds into 6 toxicity classes (1 to 6) with class 1 been the most toxic and lethal having an estimated lethal dose (LD50) of ≤ 5 while class 6 demonstrates LD50 > 5000 denoting non-toxicity of the compound. From **table 6**, LD50, organ toxicity that comprises of hepatotoxicity, carcinogenicity, and cytotoxicity of camostat, calothrixin A, apigenin, catechin, epicatechin and quercetin were predicted. The toxicity class and prediction accuracy of each compound was also displayed. All the compounds but quercetin was forecasted to be non-hepatotoxic, non-carcinogenic and non-cytotoxic with high prediction accuracy of ≥ 68.07 %. Catechin and epicatechin are especially distinguishable amongst other compounds with regards to their non-toxicity and safety as illustrated by their high LD50 of 10,000 mg/kg each and a prediction accuracy of 100 % each. The safety of these compounds surpasses that of the standard drug; camostat, which has LD50 of 3000 mg/kg with a prediction accuracy of 68.07 %. We can thus deduce that, following ADME and toxicity calculation studies, the order of the safety profile of the selected hits follow the trend; catechin and epicatechin > camostat > apigenin > calothrixin A > quercetin. This result indicates that, catechin, epicatechin and apigenin could be developed as drugs and may likely compete with the existing molecule; camostat in the inhibition of TMPRSS2 and may be used as therapeutic against SARS-COV-2.

**Table 6**. Toxicity prediction of the selected compounds.

| Compound | LD50 (mg/kg) | Toxicity | | | | | | Toxicity class | prediction accuracy |
|---|---|---|---|---|---|---|---|---|---|
| | | Hepatoxicity | | Carcinogenicity | | Cytotoxicity | | | |
| | | Prediction | Probability | Prediction | Probability | Prediction | Probability | | |
| Camostat | 3000 | inactive | 0.72 | inactive | 0.54 | inactive | 0.59 | 5 | 68.07 |
| Calothrixin A | 2000 | inactive | 0.52 | inactive | 0.51 | inactive | 0.61 | 4 | 68.07 |
| Apigenein | 2500 | inactive | 0.68 | inactive | 0.62 | inactive | 0.87 | 5 | 70.97 |
| Catechin | 10000 | inactive | 0.72 | inactive | 0.51 | inactive | 0.84 | 6 | 100 |
| Epicatechin | 10000 | inactive | 0.72 | inactive | 0.51 | inactive | 0.84 | 6 | 100 |
| Quercetin | 159 | inactive | 0.69 | active | 0.68 | inactive | 0.99 | 3 | 100 |

## 3.8 Docking interaction analysis of the top hits

The analysis of the interaction pattern of the top four lead compounds including the reference drug were considered. Although all the 17 best hits demonstrate good binding scores above or below that of the standard drug, the ADME and toxicity studies allowed us to make further screening and finally identified camostat, apigenin, catechin and epicatechin as the top lead with good pharmacokinetic and pharmacodynamic profiles. The analysis of binding interaction of these compounds with TMPRSS2 is therefore very essential as it will pave a way for their additional development as drugs with promising potency against SARS-CoV-2. The binding energies of camostat, apigenin, catechin and epicatechin obtained from molecular docking studies were -7.8 kcal/mol, -7.6 kcal/mol, -7.4 kcal/mol and -7.4 kcal/mol respectively (**table 3**).

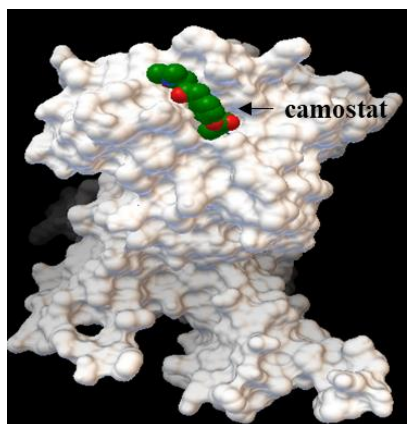

a (i)

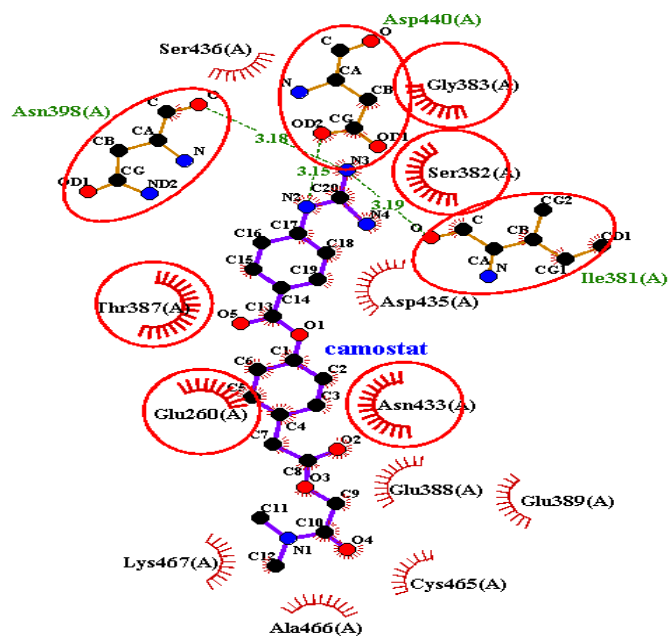

a (ii)

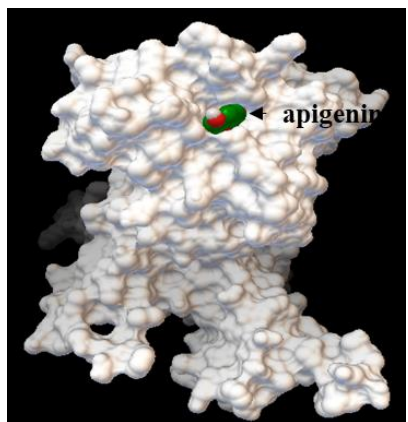

b (i)

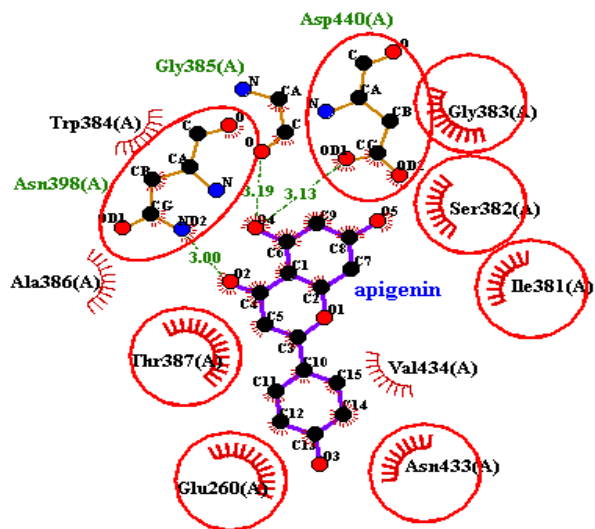

b (ii)

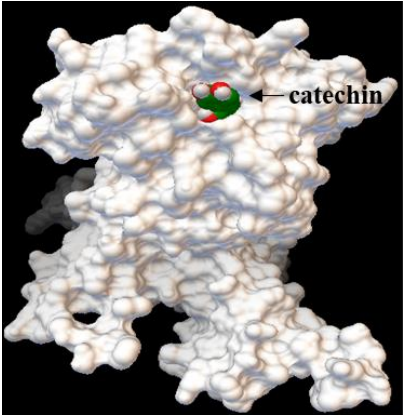

c (i)

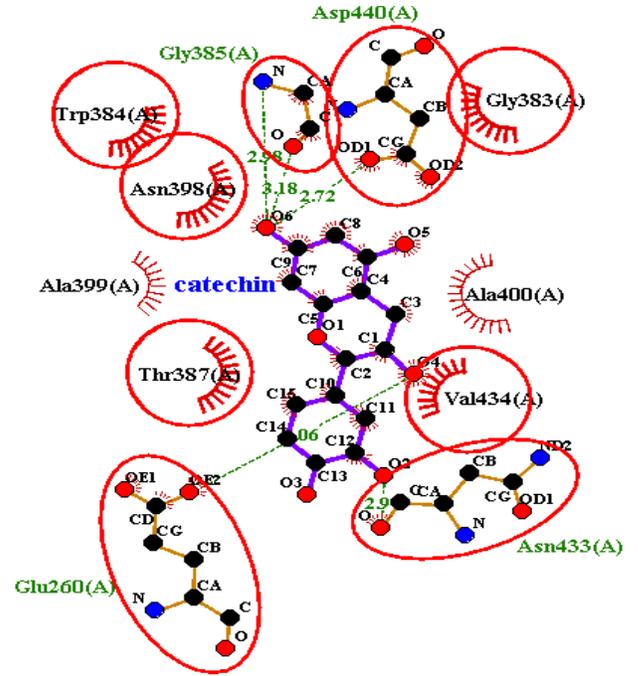

c (ii)

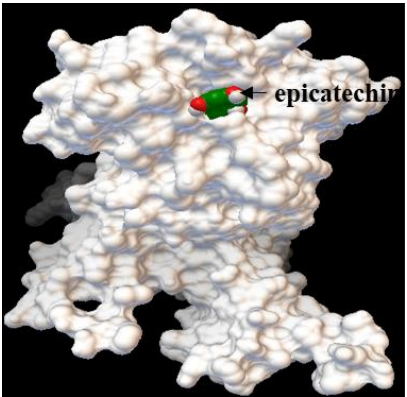

d (i)

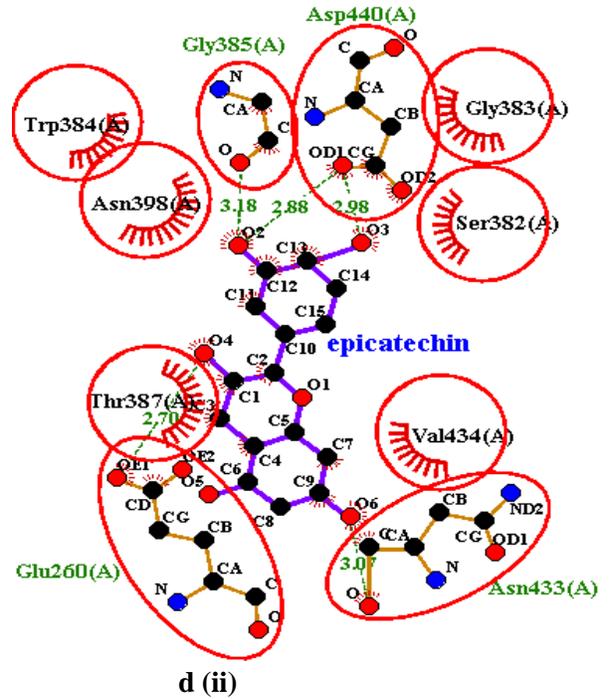

d (ii)

**Figure 6.** Binding poses (left) and 2D-interaction (right) of TMPRSS2 and the best four leads: **a (i) & ii,** TMPRSS2-camostat complex; **b (i) & ii,** TMPRSS2-apigenin complex; **c (i) & ii**, TMPRSS2-catechin complex and **d (i) &ii**, TMPRSS2-epicatechin complex.

These compounds were capable of associating withTMPRSS2 through range of contacts with different amino acid residues at virtually same binding pose of the protein (**Figure 6** (**ai-di**, left side)). These interactions are very indispensable in the formation and stabilization of docking complexes. To gain an insight into the nature and mode of interaction as well as amino acids residues involved, we used LigPlot$^+$ (version v.2.2.4). The microalgal compounds interacted with TMPRSS2 through both polar (hydrogen bonds) and non-polar (hydrophobic) contacts. The BE of camostat (-7.8 kcal/mol); a reference inhibitor drug was due to the formation of three hydrogen bonds with Asn398, Asp440, and Ile381 amino acids residues with a bond length of 3.18 Å, 3.15 Å and 3.19 Å respectively. The binding energy of the compound was also contributed by hydrophobic interaction with 12 amino acid residues of the protein which include Ser436, Gly383, Ser382, Thr387, Asp435, Glu260, Asn433, Glu388, Glu389, Cys465, Lys467, and Al466 (**Figure 6 aii**). Apigenin; a flavonoid compound found in microalgae such as *Acanthophora spicifera* [44] had a binding score of -7.6 kcal/mol derived from three hydrogen bonds with Asn398, Asp440 Gly385. These residues were positioned at nearly same distance from the ligand; apigenin with a hydrogen bond length of 3.00 Å. 3.13 Å and 3.19 Å for Asn398, Asp440 Gly385 respectively. Hydrophobic interaction with 9 amino acids namely, Trp384, Ala386, Thr387, Glu260, Gly383, Ser382, Ile381, Val434 and Asn433 were further indispensable to the BE of apigenin (**Figure 6bii**). It's quite imperative to mention that a slightly higher binding energy of camostat in comparison to that of apigenin may be attributed to more hydrophobic interaction in camostat than in apigenin. Based on the identity of the amino acids engaged in both hydrogen and hydrophobic interactions, we can also infer that both compounds bind to the same binding cavity of the protein. Catechin and its isomer epicatechin are known to be effective anticancer and anti-inflammatory agents found in Chlorella *pyrenoidosa* [45]. The binding energy of these compounds with TMPRSS2 was found to be -7.4 kcal/mol each. Their interaction with the protein was achieved through 5 hydrogen bonds with 4 amino acids residues of the protein. The residues that formed hydrogen bonds in both catechin and epicatechin are identical. These are Gly385, Asp440, Glu260 and Asn433 (**Figure 6cii & dii**). Catechin formed 2 hydrogen bonds with Gly385 with a bond distance of 2.98 Å and 3.18 Å while the proximity of the compound with Asp440, Glu260 and Asn433 via hydrogen bonds are 2.72 Å, 06 Å and 2.9 Å respectively. Contrary to catechin, epicatechin established contact with Asp440 through two hydrogen bonds within 2.98 Å and 2.88 Å and three hydrogen bonds with Gly385, Glu260 and Asn433 having a length of 3.18 Å, 2.7 Å and 3.07 Å respectively. In contrast to the number and identity of amino acids residues participated in hydrogen bonds with the compounds, catechin and epicatechin formed 7 and 6 hydrophobic contacts with the protein respectively. The hydrophobic linkage of catechin and the protein residues was established through Trp384, Asn398, Gly383, Ala399, Thr387, Val434 and Ala400 while that of epicatechin was made via Trp384, Asn398, Gly383, Ser382, Thr387 and Val434 (**Figure 6cii & dii**). From the interaction analysis, it was clearly observed that, the four top hits bind to the TMPRSS2 in the same pocket via both hydrophilic and hydrophobic linkages. The analysis of the predicted amino acids residues at the catalytic triad and substrate binding sites of TMPRSS2 have indicated that, the compounds bind closely to these sites which may induce conformational change to the protein and affect its activity. The binding energy mainly depends on the extent of this linkages with more interactions having the highest energy. Inhibiting the activity or suppression the expression of TMPRSS2 have been substantially reported to be quite reliable and efficient approach for the therapy of infections caused by viruses such as SARS-Cov-2, SARS-CoV, MERS-CoV, and influenza viruses that utilizes TMPRSS2 for entry into the host cell [46]–[48].

## 4. Conclusion

Owing to the severity and paucity of efficient drugs, the novel coronavirus become a global health burden of emergency concern. Research on the development of effective treatment strategies is inevitably necessary. Studies of protein-small organic molecules interaction is a powerful method in drugs development. This process is highly laborious, time demanding and costly in the laboratory. In silico or computational approach was found highly promising in the simplification of drugs design and discovery. In this study, an in-silico method was used to screen 95 natural compounds from microalgae with a view to identify the best compounds capable of binding and inhibiting TMPRSS2 protein implicated in the pathogenesis of Covid-19. Molecular docking studies, ADME and Toxicity analyses led to the identification of camostat, apigenin, catechin and epicatechin as good drug candidates against TMPRSS2. These compounds exhibit good binding scores of -7.8 kcal/mol, -7.6 kcal/mol, -7.4 kcal/mol and -7.4 kcal/mol each for camostat, apigenin, catechin and epicatechin respectively. The compounds also demonstrate good pharmacokinetic and pharmacodynamic profiles based on ADME and Toxicity studies. Overall result shows that all the compounds fit in the same binding cavity of the protein and may serves as suitable drug candidates against Covid-19. However, further in vitro and in vivo studies of the identified compounds need to be carried out prior to their further stage of development.